# The Human-AI Handshake Framework: A Bidirectional Approach to Human-AI Collaboration

Aung Pyae

International School of Engineering, Faculty of Engineering, Chulalongkorn University

aung.p@chula.ac.th

Abstract: Human-AI collaboration is evolving from a tool-based perspective to a partnership model where AI systems complement and enhance human capabilities. Traditional approaches often limit AI to a supportive role, missing the potential for reciprocal relationships where both human and AI inputs contribute to shared goals. Although Human-Centered AI (HcAI) frameworks emphasize transparency, ethics, and user experience, they often lack mechanisms for genuine, dynamic collaboration. The "Human-AI Handshake Model" addresses this gap by introducing a bi-directional, adaptive framework with five key attributes—information exchange, mutual learning, validation, feedback, and mutual capability augmentation—that foster balanced interaction. This model enables AI to act as a responsive partner, evolving with users over time. Human enablers like user experience and trust, alongside AI enablers such as explainability and responsibility, facilitate this collaboration, while shared values of ethics and co-evolution ensure sustainable growth. Distinct from existing frameworks, this model is reflected in tools like GitHub Copilot and ChatGPT, which support bi-directional learning and transparency. Challenges remain, including maintaining ethical standards and ensuring effective user oversight. Future research will explore these challenges, aiming to create a truly collaborative human-AI partnership that leverages the strengths of both to achieve outcomes beyond what either could accomplish alone.

## Introduction

Recent advancements in artificial intelligence (AI) have redirected discussions from concerns about job displacement to the exploration of opportunities for human-AI collaboration. While AI automation continues to play a critical role in specific tasks, such as email spam filtering, increasing evidence from both academic research and industry practices highlights the potential of AI not only to enhance human capabilities but also to facilitate the emergence of novel forms of work that integrate human expertise with computational strengths. For instance, in healthcare, AI-assisted decision-making systems integrated with advanced machine learning models can enhance diagnostic accuracy by analyzing complex patient data that is challenging or impossible for humans to detect (Cai et al., 2019; Lai et al., 2021). In the creative domain, AI-driven systems can collaborate with artists and designers in processes such as ideation, prototyping, and iteration, enhancing human artistic capabilities while preserving creative originality (Rezwana & Maher, 2023). These examples illustrate the potential of AI to augment human capabilities rather than replace human roles and responsibilities.

Over the past decade, researchers and practitioners across various disciplines have reached a consensus that the value of human-AI collaboration lies in effective design and dynamic interactions that amplify both human agency and AI capabilities. Existing literature has introduced well-established models and frameworks for optimizing human-AI interaction and collaboration. For instance, the Co-Creative Framework for Interaction Design (COFI) by Rezwana and Maher (2023) identifies interaction spaces within co-creative systems. Wang and Yin (2021) emphasize that transparent explanations provided by AI are critical for fostering trust in human-AI interactions. Similarly, Beghetto (2023) underscores the importance of preserving human agency in creative processes to ensure that AI serves as a complement to, rather than a replacement for, human efforts and capabilities. Collectively, these studies highlight that well-designed collaborative AI systems should enhance human creativity and abilities while upholding principles of transparency, accountability, trust, and user experience.

Bi-directional communication and interaction have increasingly been recognized as essential for advancing human-AI collaboration, particularly as AI transitions from passive tools to active collaborators. This shift enables AI to augment human capabilities and facilitates more meaningful interactions to achieve shared goals or tasks. For instance, GitHub Copilot, a widely used AI-assisted tool for software development, provides code suggestions while developers validate and provide feedback, forming a dynamic loop where both human and AI contribute meaningfully to a common objective. However, this does not imply that AI is equivalent to humans in task performance. Instead, it highlights AI's computational power and adaptability in complementing human abilities. In the academic literature, Jiang et al. (2022) emphasize the need to enhance AI's explainability to empower user influence, while Cai et al. (2019) advocate for 'collaborative mental models,' which enable humans and AI to leverage each other's strengths. Additionally, Ezer et al. (2019) explore 'trust engineering' as a strategy to develop effective AI-human teams by promoting transparency and adaptability. Collectively, these studies underscore the pivotal role of two-way communication and interaction in fostering effective human-AI collaboration.

Existing research often frames AI in a subordinate role within human-AI interactions, which can undervalue its potential as an interactive collaborator and complementary partner. Additionally, the current body of literature lacks sufficient emphasis on bi-directional relationships in human-AI collaboration and the attributes necessary for such partnerships. Furthermore, there is limited understanding in existing research regarding the enablers critical to facilitating effective human-AI collaboration. According to Yue (2023) and Jiang et al. (2022), effective human-AI collaboration should prioritize user needs, uphold user autonomy, and leverage AI's adaptability. However, these Human-Centered AI (HCAI) principles remain underexplored in the development of contemporary theoretical models and technical frameworks.

Given the growing body of research on AI as a complementary partner to human capabilities and the increasing interest in bi-directional interactions within human-AI collaboration, it is imperative to reconceptualize AI as a dynamic, interactive partner that enhances and augments human abilities while safeguarding human agency. Furthermore, ethical standards must be upheld throughout this collaborative process. Addressing these existing gaps and integrating Human-Centered AI (HCAI) principles—such as trust, user experience, explainability, and adaptability—into AI systems can unlock their full potential, fostering more effective, trustworthy, and ethically aligned human-AI collaborations.

This study addresses existing research gaps by proposing the human-AI handshake framework, which emphasizes bi-directional collaboration. The framework identifies key bi-directional attributes, such as information exchange, and introduces enablers, including user experience, ethics, and explainability, which are critical for fostering effective human-AI collaboration. The handshake metaphor conceptualizes productive collaboration by positioning AI as a complementary partner that enhances human abilities while ensuring that responsibility remains human-led. The framework is grounded in an extensive review of the literature on human-centered AI, human-AI collaboration, and user-centered design, setting it apart by prioritizing bi-directional interaction. Furthermore, a review of existing AI tools demonstrates alignment with the proposed handshake framework while highlighting their limitations. These insights inform a research agenda that incorporates expert interviews and user feedback to refine and validate the framework. This project seeks to advance the field of human-AI cooperation and interaction by conceptualizing AI as a dynamic, partner-like tool through the human-AI handshake framework. It aims to establish new possibilities for collaboration that are not only reliable and effective but also ethically grounded.

The study aims to:

- Introduce the human-AI handshake framework, detailing attributes and enablers that foster bi-directional collaboration.
- Investigate how AI can act as a complementary partner, enhancing human abilities while upholding responsibility and ethical standards.
- Develop a research agenda to refine the framework through interviews and surveys with users and experts, identifying applications and challenges.

## Literature Review

### *Human-Centered AI (HCAI)*

In recent years, human-centered artificial intelligence (HCAI) represents an important paradigm shift in AI design and human-computer interaction that focuses on user experience, ethical considerations, and collaboration between humans and AI. One of the core principles of HCAI is the focus on human-centered design, user perceptions, and cognitive processes in AI design and development. For instance, Shin (2021) highlights that understanding user cognition plays an important role in developing algorithmic services that are user-centered and rational, ultimately fostering better user experiences, trust, and acceptance. Similarly, Gomaa (2022) describes the growing interest in HCAI among researchers and practitioners, particularly due to the urgent need for reliable, explainable, and trusted interfaces that enhance user engagement as well as user experience.

Incorporating human-centered principles is crucial for enhancing trust and usability in AI systems, as echoed by Shneiderman (2020), who advocates for integrating human considerations to promote reliability, user experience, and user satisfaction. In addition, ethical considerations play a critical role in HCAI, as they ensure responsible AI design and development with a focus on societal impact. This is enriched by Rezwana and Maher (2022)'s study in which they argue that technological capabilities alone do not guarantee positive collaboration; instead, AI must enhance rather than replace human capabilities. Shneiderman (2020) also advocates for balancing human control with automation to optimize safety and performance in HCAI. These perspectives in HCAI highlight that ethical frameworks are crucial for ensuring AI systems are beneficial and aligned with human values and needs.

Another important component of HCAI is explainability. According to Liao et al. (2020), user-centered approach plays a critical role to explainable AI (XAI) that addresses users' questions and concerns, thereby improving user experience, trust, and satisfaction, fostering better human-AI collaboration. IIt is supported by Miller (2019) who suggests that social science insights can inform AI explanations, making AI systems more transparent and explainable, and enhancing user interactions and trust. These studies clearly underscore the importance of XAI essential for bridging the gap between AI capabilities, and user understanding and trust. It can also ensure that AI technologies are approachable and trustworthy. The collaborative nature of human-AI interactions is another important theme in HCAI. This claim is supported by the study by Rezwana and Maher (2022) in which the researchers indicate that effective communication and interaction between AI systems and users can lead to greater perceived reliability and intelligence, particularly in co-creative contexts. This underscores the importance of designing human-centered AI systems that engage users meaningfully and collaboratively rather than merely performing tasks. Such interactions can enhance user experiences and establish AI as a collaborative partner rather than just a tool.

In summary, the existing literature on HCAI underscores the importance of integrating user perspectives, ethical considerations, and collaborative frameworks into AI design and implementation. A human-centered approach is vital to ensure that AI technologies are effective, trustworthy, and aligned with user needs and human values, especially as AI continues to become more deeply embedded in everyday life.

### *Human-AI Collaboration*

Recently, in both academic and industry, there is a growing interest in human-AI collaboration that leverages technology to enhance human capabilities across various fields including healthcare and education. This method goes beyond traditional ideas of AI as just a tool and focuses on a synergistic interaction and communication in which humans and AI systems collaborate to achieve common goals and tasks. For example, in healthcare, AI-driven systems assist clinicians in diagnosing diseases by analyzing complex datasets, while clinicians contribute contextual knowledge and feedback to refine treatment strategies (Järvelä et al., 2023). This dynamic interaction and collaborative nature exemplifies how human expertise and AI's computational abilities and advantages can combine to deliver more effective outcomes.

The literature highlights that the effectiveness of human-AI collaborations relies on critical factors such as trust, ethics, transparency, and user experience. For instance, Agbese et al., (2021) highlight that trust is fundamental in such collaboration, as it influences how much users rely on AI-generated insights and recommendations. Furthermore, ethical frameworks are also crucial, ensuring that AI systems operate with fairness, accountability, and inclusivity, thereby fostering a collaboration and interaction that human and AI capabilities complement each other effectively (Rezwana & Maher, 2022). Also, transparency in AI systems, alongside a positive user experience, is essential for building user confidence, trust, and enabling smooth integration of AI systems into human workflows.

In recent years, human-AI collaboration has been applied in various sectors. For instance, in creative industries, commercial tools like DALL-E and ChatGPT enable artists, writers, content creators to generate novel concepts and innovative contents, blending AI's generative capabilities with human originality and artistic values (Rezwana & Maher, 2023). In business, AI solutions such as Microsoft's Copilot and Salesforce's Einstein offer the ability to automate routine tasks and generate insights, empowering professionals to focus on strategic decision-making (Fu, 2023). In healthcare, AI systems like IBM Watson has the power to analyze extensive medical datasets to recommend treatments, while clinicians tailor these recommendations to individual patient needs (Järvelä et al., 2023). In transportation, autonomous technologies developed by Tesla and Waymo rely on human oversight to enhance safety and efficiency, illustrating the potential of hybrid control systems (Dikshit, 2023). These applications show how human-AI collaborations might be revolutionary, but they also highlight that more work is required to achieve true flexibility and co-evolution in human-AI collaobration.

### *Research Gaps*

Despite advancements, significant gaps persist in achieving effective human-AI collaboration, particularly in establishing robust bi-directional relationships where AI systems dynamically adapt to user needs and meaningfully incorporate human feedback. Many existing AI systems prioritize task efficiency over fostering meaningful engagement, often failing to act as truly complementary and adaptive partners. This shortfall results in misalignments with user expectations and inadequately addresses human values, autonomy, and decision-making authority. Moreover, current designs frequently lack mechanisms to balance automation with transparency and accountability, reducing user trust and engagement.

The concept of "bi-directional" interaction, where humans and AI actively influence and adapt to one another in real time, remains underexplored. Existing systems often lack the capacity for truly fluid and responsive partnerships, underscoring the need for further research into adaptive AI systems capable of continuous learning and mutual feedback. Addressing these gaps is critical for advancing the field and achieving more effective, trustworthy, and ethically grounded human-AI collaborations (Ashktorab et al., 2020).

To address these limitations, a paradigm shift toward collaborative, human-centered frameworks is essential. Such frameworks must emphasize user-centered design, responsibility, and adaptability while embedding ethical considerations and prioritizing user experience. AI must function as a partner-like tool, augmenting human capabilities while humans retain full responsibility for decisions and outcomes. Both human and AI collaboration must respect ethical principles to ensure that this partnership remains trustworthy, accountable, and aligned with societal values. By transitioning AI from a static tool to a dynamic and supportive collaborator, human-AI collaboration can better enhance human activities while maintaining human oversight and ethical integrity.

## Human-AI Handshake Framework

The human-AI handshake framework was developed based on a comprehensive literature review of foundational works in HCAI, HCI, human-centered AI design, and human-AI collaboration. This review highlighted key themes such as human-centered design, ethical frameworks, bi-directional relationships, and collaborative dynamics (Agbese et al., 2021; Rezwana & Maher, 2022; Shneiderman, 2020). Principles from these studies emphasized the importance of fostering bi-directional interactions where AI dynamically augments human capabilities while humans retain accountability and ensure ethical alignment. To refine the framework, feedback was incorporated from AI researchers and practitioners. Their insights helped address practical challenges, ensuring the framework aligns with real-world needs and supports collaboration through adaptability, explainability, and trust. These contributions further shaped the framework to emphasize information exchange, mutual learning, and feedback, creating a foundation for dynamic human-AI partnerships that prioritize usability and ethical considerations. The final framework synthesizes findings from the literature with expert feedback to provide a cohesive model for promoting effective, trustworthy, and human-centered AI collaboration across diverse applications. Figure 1 illustrates the human-AI handshake framework, showing how its components work cohesively to address user needs, promote trust, and ensure collaborative efficiency. By embedding ethical principles, explainability, adaptability, and reliability, the framework provides a robust model for fostering effective, accountable, and ethically aligned human-AI partnerships.

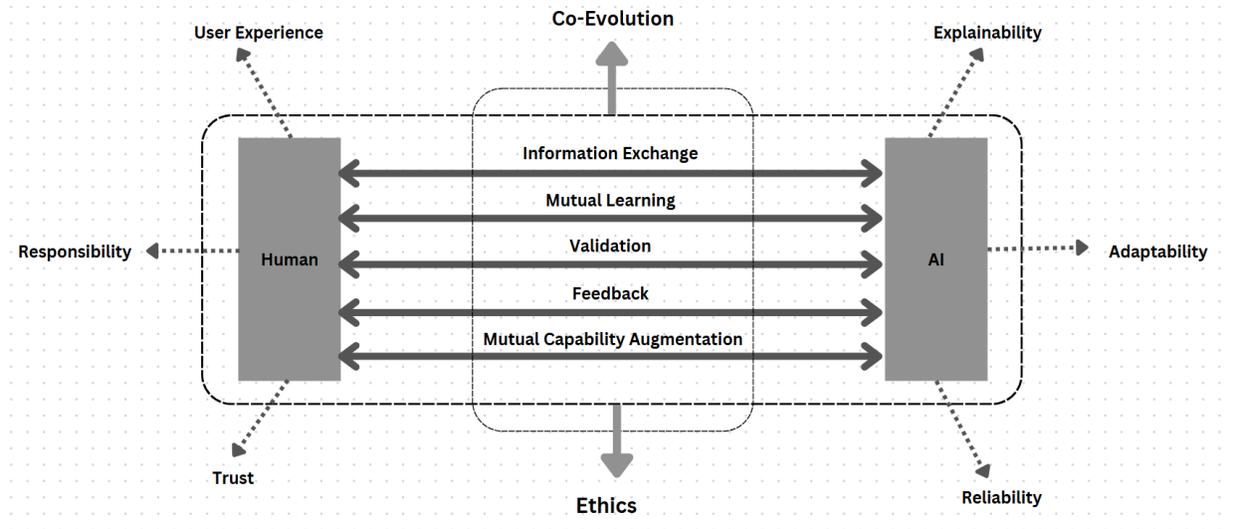

**Figure 1. Human-AI handshake Model**

## *Handshake Metaphor and Bi-directional Relationship*

The handshake metaphor in this framework symbolizes a collaborative relationship between humans and AI, highlighting adaptive support and mutual reinforcement toward shared goals. AI functions as a partner-like tool, augmenting human capabilities while respecting human values, without assuming equal responsibility or shared accountability. AI's role is to be responsive and adaptable, enhancing human potential while ethical responsibility and decision-making authority remain with the human user. For instance, in a personalized learning platform, AI tailors instructional content based on student feedback, providing adaptive support while the student retains full control over the learning process and guides the AI's adjustments. This illustrates the handshake concept, where AI serves as an ethical and adaptive collaborator that enhances outcomes while humans maintain accountability and agency. The bi-directional relationship in this framework fosters a dynamic partnership where humans and AI exchange information, learn, and adapt together to achieve shared goals. While AI provides adaptive support to enhance decision-making, efficiency, and capabilities, humans retain full control and accountability. This collaboration ensures mutual reinforcement and ethical co-evolution, with AI amplifying human values and capabilities while respecting the human's ultimate authority.

## *Bi-directional Attributes*

### Information Exchange

Bi-directional information exchange is central to the human-AI handshake framework, enabling adaptive, two-way communication where AI functions as a partner-like tool responsive to human needs. This dynamic interaction allows AI to receive contextual feedback while providing humans with knowledge and outputs that align with their expectations, ultimately enhancing decision quality (Taudien et al., 2022). By incorporating non-verbal cues, contextual signals, and ongoing feedback, AI can better interpret human needs while offering transparency in its decision-making process. This exchange fosters seamless collaboration, ensuring human control, accountability, and autonomy. Transparency is key, as it builds trust, user satisfaction, and engagement, while its absence hinders acceptance (Wang & Yin, 2021). Though challenges remain, prioritizing user-centered design with transparency, explainability, and adaptability is crucial to bridge gaps between AI capabilities and human needs (Rezwana & Maher, 2023). In fields like architectural design, this reciprocal flow allows AI to propose innovative solutions while human input ensures alignment with contextual and aesthetic values. Humans benefit from AI's creativity and analysis, reinforcing mutual adaptability and ensuring AI augments human potential while responsibility remains with the human (Petráková, 2023).

**Mutual Learning**

Bi-directional mutual learning is central to the human-AI handshake framework, emphasizing a continuous, interactive process where AI acts as a partner-like tool that augments human capabilities. Through shared experiences, reciprocal feedback, and adaptation, both humans and AI enhance their skills while humans retain ultimate control and accountability. AI's computational power complements human cognition, fostering a co-evolutionary learning process that improves outcomes in fields like education and healthcare (Järvelä et al., 2023). Mutual learning enables AI to adapt to human preferences and behaviors, while humans benefit from AI's analytical strengths, enriching collaboration without implying equal roles or shared accountability. In educational settings, AI provides personalized feedback and adapts to individual learning styles, enhancing learning effectiveness while educators maintain responsibility for outcomes (Akavova, 2023). Similarly, Nguyen et al. (2022) illustrate how multimodal learning analytics offer timely feedback, boosting collaborative success without diminishing human oversight. Beyond education, mutual learning extends to organizational environments where AI augments human capabilities to enhance innovation and job performance, with humans retaining decision-making authority (Chen et al., 2023). Jarrahi et al. (2022) argue that AI's potential is realized through reciprocal learning, stressing the importance of engaging AI systems to learn actively from human users without equating their roles. This bi-directional learning fosters trust, adaptability, and alignment, supporting the human-AI handshake framework's goal of dynamic adaptability and mutual reinforcement. Ultimately, mutual learning ensures that AI evolves alongside human partners to bridge the gap between AI capabilities and human needs, enhancing collaboration while maintaining human oversight and ethical responsibility.

**Validation**

Bi-directional validation is fundamental to the human-AI handshake framework, highlighting AI as a partner-like tool that augments human capabilities through mutual verification and adaptation. This continuous feedback loop allows AI to validate human inputs while enabling humans to verify AI outputs, fostering trust essential for adaptive learning (Rezwana & Maher, 2023). This process enhances mutual learning by promoting transparency—defined as a mutual understanding of actions and intentions—which addresses the "black box" issue that can erode user trust (Holder et al., 2021; Sawant, 2023). Effective validation mechanisms are crucial in fields like healthcare, where humans confirm AI-generated diagnoses to ensure patient safety (Lai et al., 2021), and in organizational contexts, where verifying AI-driven insights enhances strategic decision-making without diminishing human judgment (Alstete & Meyer, 2020). Governance frameworks that promote mutual validation strengthen ethical and reliable collaboration, reinforcing human oversight and trustworthiness (Garibay et al., 2023; Sigfrids et al., 2023). By integrating advanced explainability and real-time feedback, bi-directional validation supports dynamic human-AI partnerships, aligning AI capabilities with human needs and emphasizing mutual learning to build trust, adaptability, and reliable collaboration.

**Feedback**

Bi-directional feedback is fundamental to the human-AI handshake framework, emphasizing a collaborative partnership where AI acts as a tool that augments human capabilities through continuous learning and adaptation. This reciprocal process enhances decision-making and performance outcomes by fostering mutual learning between humans and AI systems. Akavova (2023) highlights this in educational contexts, where adaptive learning systems provide personalized feedback to students while simultaneously refining their algorithms based on student interactions. Effective feedback hinges on the design of human-AI collaborative systems; Jain et al. (2022) note that tailored, contextually relevant feedback enhances collaboration, especially when AI complements rather than replaces human expertise. In healthcare, bi-directional feedback improves patient outcomes by allowing professionals to refine AI algorithms and leverage AI insights for clinical decisions (Lai et al., 2021). Holder et al. (2021) underscore the role of bi-directional transparency in enabling feedback loops by clarifying actions and intentions, while Shneiderman (2020) observes that reciprocal feedback empowers iterative refinement of both human and AI behaviors, ensuring relevance in real-world applications. Despite its benefits, further research is needed to expand our understanding of bi-directional feedback across domains like healthcare and education, ensuring that AI systems evolve as collaborative partners to enhance mutual trust and performance outcomes.

**Capability Augmentation**

Bi-directional capability augmentation is fundamental to the human-AI handshake framework, emphasizing a synergistic partnership where AI acts as a partner-like tool that augments human capabilities through mutual learning and adaptation. This bi-directional relationship leverages the unique strengths of both humans and AI to overcome limitations, resulting in outcomes greater than the sum of their parts across fields like education, healthcare, and project management. In K-12 education, AI personalizes learning and provides data-driven insights, while educators contribute relational context and emotional intelligence, leading to improved educational outcomes (Holstein & Aleven, 2022). In project management, AI's data analysis and pattern recognition complement human judgment and creativity, enabling nuanced risk management (Nyqvist, 2024). In healthcare, AI analyzes medical data to inform clinical decisions, while practitioners validate and apply these insights for patient-centric care (Lai et al., 2021). Combining AI's data-processing strengths with human expertise elevates decision quality and care standards (Bailer et al., 2022; Klumpp, 2017). Transparency and trust are essential in this mutual augmentation; reasoned transparency clarifies AI processes and roles (Carrubbo, 2024), and the dynamic learning and adaptation of both agents enhance this synergistic partnership (Jain et al., 2022). By integrating complementary roles and fostering transparent interactions, bi-directional capability augmentation enhances decision-making, builds trust, and revolutionizes human-AI collaboration, though further research is needed to optimize these dynamics.

## *Enablers for Human-AI Collaboration*

**Enablers for Human in Human-AI Collaboration**

Enablers on the human side are crucial for effective human-AI collaboration within the human-AI handshake framework. Key among these are user experience (UX), trust, and user responsibility, which collectively ensure a bi-directional partnership that enhances engagement, understanding, and decision-making between humans and AI systems. A positive UX makes AI systems intuitive, accessible, and responsive, aligning interactions with human cognitive capabilities to minimize friction and complexity. Adaptability and personalization in UX allow users to easily understand and navigate AI functionalities, fostering comfort and encouraging valuable feedback that strengthens the human-AI partnership. Trust is essential for integrating AI into decision-making processes; users must trust the reliability, accuracy, and transparency of AI systems to rely on them effectively. This trust is built through consistent, explainable AI behavior and mechanisms like human validation of AI outputs, which reinforce user confidence and system accuracy (Yang et al., 2020; Sreedharan, 2023). User responsibility is also vital; users need to actively engage with AI systems, understand their capabilities and limitations, and maintain an accurate mental model of their functionality (Rezwana & Maher, 2023; Dhuliawala, 2023). Effective collaboration depends on users calibrating their trust appropriately and providing feedback to improve the system while being aware of their own cognitive biases (Okamura & Yamada, 2020; Naiseh et al., 2023; Rastogi et al., 2022). By emphasizing these enablers—user experience, trust, and user responsibility—the human-AI handshake framework fosters AI systems that are user-friendly, reliable, and adaptable, ultimately enabling smoother and more effective bi-directional collaboration.

**Enablers for AI in Human-AI Collaboration**

Enablers on the AI side are crucial for effective human-AI collaboration within the human-AI handshake framework. Key among these are explainability, reliability, and adaptability, which collectively drive transparency, accountability, and human trust in AI systems. Explainability is essential for creating AI systems that are not perceived as "black boxes"; AI must articulate the reasoning behind its decisions in a way humans can understand, fostering trust and enabling informed decision-making, especially in high-stakes domains like healthcare, finance, and law (Sharma et al., 2022). Reliability influences the effectiveness of these partnerships; while AI can enhance human performance, it often struggles with complex, nuanced interactions, leading users to question its dependability, particularly in critical environments like healthcare (Doraiswamy et al., 2020; Asan et al., 2020). Calibration of trust is essential to avoid over-reliance or under-utilization of AI, which can compromise decision-making quality (Duarte, 2023; Okamura & Yamada, 2020). Adaptability is pivotal for AI systems to evolve based on shared experiences and feedback, adjusting to the nuances of human behavior and contexts to provide relevant support (Puranam, 2020; Rezwana & Maher, 2023). By emphasizing these enablers—explainability, reliability, and adaptability—the human-AI handshake framework fosters AI systems that are transparent, trustworthy, and capable of evolving alongside human partners, ultimately enhancing collaboration and outcomes.

### *For Both Users and AI*

Co-evolution and ethics are shared enablers in the human-AI handshake framework that foster effective collaboration between humans and AI. Co-evolution refers to the reciprocal development of human and AI capabilities, where both parties adapt and learn from each other. AI systems refine their algorithms based on human input to better align with user needs, while humans enhance their ability to leverage AI by refining decision-making and integrating AI into workflows. This mutual evolution creates a collaborative environment where both advance together. Madni (2020) explores augmented intelligence, framing AI as a partner that amplifies human abilities to achieve outcomes unattainable alone, highlighting the necessity of evolving together. Similarly, Holder et al. (2021) emphasize bi-directional transparency in human-AI-robot teams, fostering mutual understanding and ongoing learning, which enhances collaboration.

Ethics serves as another essential enabler, guiding interactions to ensure fairness, accountability, and alignment with shared values. Ethical principles shape AI design and deployment by emphasizing fairness, transparency, accountability, and respect for privacy. AI systems aim to minimize biases and uphold privacy, while humans are encouraged to use AI responsibly, considering societal implications. Raisch and Krakowski (2020) discuss the automation-augmentation paradox, advocating for augmentation over automation to prioritize collaborative outcomes, aligning with ethical principles by valuing contributions of both humans and AI. Furthermore, Alevizos (2024) addresses the ethical implications of AI in sensitive areas like cyber threat intelligence, underscoring the importance of ethical frameworks in guiding AI applications that impact privacy and security, thereby fostering trust in AI-human partnerships. By emphasizing co-evolution and ethics within the human-AI handshake framework, organizations can foster productive and responsible collaboration. Prioritizing mutual growth, transparency, and shared responsibility allows both humans and AI to leverage their strengths, driving positive outcomes and contributing to societal well-being.

### *Review of the Human-AI Handshake Framework in the Context of Existing AI Tools*

This review evaluates how current AI tools—GitHub Copilot, ChatGPT, Figma, Adobe AI, and conceptual AI agents—align with the human-AI handshake framework. It examines their strengths and limitations in fostering dynamic, bi-directional collaboration, focusing on attributes like information exchange, mutual learning, user experience, and explainability. By assessing these tools, which span fields like software development, content creation, and design, the review provides insights into how well they support synergistic human-AI partnerships and identifies areas for improvement to enhance integration across diverse domains.

**A Case of GitHub Copilot**

GitHub Copilot strongly aligns with the human-AI handshake framework by supporting bi-directional collaboration through information exchange, feedback, validation, mutual learning, and capability augmentation. It excels in providing context-aware, real-time code suggestions based on developer input, enhancing productivity. However, its effectiveness can diminish in complex scenarios where it misinterprets context, leading to less precise suggestions. Feedback mechanisms are naturally embedded, allowing developers to accept, reject, or modify suggestions. While this interaction enables refinement, Copilot lacks explicit learning from these corrections, limiting its real-time adaptability. Validation is upheld as developers review and approve AI-generated code, ensuring human accountability. Yet, limited explainability hinders transparency and complicates trust in critical applications; enhancing recommendations with clear explanations would strengthen reliability. Mutual learning occurs as Copilot adapts to individual coding styles over time, promoting a symbiotic relationship between tool and user. However, reliance on static training data restricts its ability to dynamically respond to evolving project requirements. Capability augmentation is evident in how Copilot automates repetitive tasks and fosters creativity, but its lack of ethical oversight could limit utility in sensitive scenarios.

From a user perspective, Copilot enhances the experience through seamless integration into development environments and an intuitive interface, streamlining workflows and improving satisfaction. Developers retain control over reviewing and approving code suggestions, reinforcing accountability. Trustworthiness is undermined by occasional contextual inaccuracies and a lack of explainability, which can reduce confidence in outputs for complex tasks. Regarding co-evolution and ethics, Copilot's static training data limits its ability to evolve alongside changing project needs, constraining true co-evolution. It lacks mechanisms for ethical oversight, such as flagging potentially harmful code or adhering to compliance standards, highlighting areas for improvement.

From an AI perspective, Copilot's lack of detailed explanations for code suggestions reduces transparency, making it harder for developers to fully trust its outputs. While it adapts to individual coding preferences, it may struggle with novel or rapidly changing requirements due to static training data. Reliability is strong for routine tasks but may falter in complex or high-stakes scenarios because of contextual limitations. In summary, GitHub Copilot showcases key strengths in aligning with the human-AI handshake framework by facilitating bi-directional collaboration and enhancing user experience and productivity. Addressing shortcomings in explainability, adaptability, and ethical oversight would bolster its effectiveness, fostering a more robust and symbiotic relationship between developers and AI..

**A Case of ChatGPT**

ChatGPT exemplifies the human-AI handshake framework by enabling bi-directional collaboration through information exchange, feedback, validation, mutual learning, and capability augmentation. It excels at providing coherent, context-aware responses, enhancing efficient communication. Users can refine queries and shape the conversation, but in complex or specific scenarios, its responses may lack depth or precision, requiring iterative prompts for clarity. While ChatGPT adapts within sessions, its reliance on static training data limits dynamic learning over time. It augments user capabilities by automating content generation and simplifying complex concepts, boosting productivity and creativity. However, it requires human oversight to validate the accuracy and relevance of outputs, especially in high-stakes applications. The lack of embedded ethical safeguards and contextual awareness may limit its reliability in sensitive or specialized contexts. From a user perspective, the intuitive conversational interface enhances the experience, but trust can be undermined by occasional factual inaccuracies or vague responses, highlighting the need for improved transparency and reliability. ChatGPT's limited explainability—failing to clarify its reasoning processes—can reduce trust in critical applications. While reliable for general tasks, it may falter in complex or domain-specific scenarios. In summary, ChatGPT aligns with the human-AI handshake framework by fostering collaboration and enhancing productivity. Addressing gaps in ethical safeguards, contextual awareness, and dynamic learning capabilities would strengthen its reliability as a robust partner in human-AI collaboration..

**A Case of Adobe AI Tools**

Adobe's AI-powered tools—such as Adobe Sensei, Photoshop's generative fill, and Premiere Pro's AI editing features—strongly align with the human-AI handshake framework by fostering bi-directional collaboration, enhancing creativity, and automating complex design and editing tasks. They excel in information exchange by providing context-aware suggestions and interpreting user intent through natural language prompts or specific selections. For example, Photoshop's generative fill allows users to modify images efficiently, and Premiere Pro's AI automates repetitive editing tasks like scene detection and noise reduction while enabling user adjustments. Feedback mechanisms are embedded, allowing users to iteratively refine AI-generated outputs within a session.

However, these tools lack dynamic learning from feedback across sessions, limiting long-term adaptability and mutual learning. Users are responsible for validating the accuracy and relevance of AI outputs, maintaining human accountability. Limited explainability of how specific edits are generated can pose challenges in creative or ethical decision-making, especially in high-stakes professional projects. The tools significantly enhance capability augmentation by automating labor-intensive tasks such as background removal, color correction, and audio leveling, allowing users to focus on higher-level creative decisions. Despite this, the potential for misuse—like generating misleading content or deepfakes—highlights the need for robust ethical safeguards. Designed with a strong emphasis on user experience, Adobe's AI tools provide intuitive interfaces that integrate seamlessly into existing workflows, democratizing advanced design and editing capabilities. While users retain responsibility for curating and approving outputs, occasional inaccuracies or less polished results may impact trust, necessitating further refinement.

Although the tools offer customizable presets and adapt to user preferences during sessions, their reliance on static training data limits dynamic adaptation to evolving user needs or creative trends. Ethical considerations, such as addressing biases and preventing unethical use of generated assets, require additional attention despite existing safeguards like content restrictions. From an AI perspective, the tools demonstrate strong adaptability within individual projects but lack long-term learning capabilities. Limited explainability can hinder trust in complex projects, and while reliability is generally high for routine tasks, it may diminish in nuanced applications requiring user intervention.

In summary, Adobe's AI tools effectively align with the human-AI handshake framework by enhancing creativity, productivity, and user experience through bi-directional collaboration. Addressing gaps in explainability, adaptability, and ethical safeguards is crucial to strengthening their reliability and trustworthiness, highlighting the importance of continuous refinement to meet evolving user expectations and ethical considerations.

**A Case of Other AI Tools**

AI-powered tools like Figma, Grammarly, Notebook LM, and scite.AI align with the human-AI handshake framework by fostering bi-directional collaboration. They excel in information exchange, offering context-aware suggestions and enabling users to refine inputs iteratively. Feedback mechanisms are embedded as users shape AI-generated content, but a lack of dynamic learning limits their adaptability and mutual learning over time. Validation and accountability are crucial, requiring users to verify the relevance and accuracy of outputs. While Grammarly refines grammar and tone, it may struggle with nuanced contexts. Figma enhances creative workflows but needs designers to validate AI recommendations. Limited explainability in these tools can hinder trust, especially in high-stakes applications. These tools augment user capabilities by automating repetitive tasks, allowing users to focus on higher-level decisions. However, the absence of robust ethical safeguards presents challenges in sensitive areas like academic research and professional design. Their static training models constrain co-evolution with users; they do not dynamically adapt to individual styles or evolving needs. Ethical considerations, such as preventing biases and ensuring integrity, remain under-addressed despite basic safeguards. From an AI perspective, they demonstrate adaptability within sessions but lack long-term learning for user alignment. Limited explainability reduces transparency, and reliability diminishes in specialized applications requiring nuanced understanding. To fully align with the human-AI handshake framework, these tools need to address gaps in explainability, adaptability, and ethical safeguards to ensure more trustworthy and effective human-AI collaboration.

In summary, the evaluated AI tools align partially with the human-AI handshake framework, excelling in information exchange, feedback, and capability augmentation. Tools like GitHub Copilot, ChatGPT, and Adobe AI enhance productivity and user experience by automating tasks and enabling bi-directional collaboration. However, their reliance on static training models limits adaptability and co-evolution, while inconsistent validation and limited explainability undermine trust, particularly in critical applications. Addressing these gaps—through dynamic learning, enhanced transparency, and robust ethical safeguards—can strengthen trust and reliability. By focusing on user experience, accountability, and adaptability, developers can create tools that foster robust, trustworthy, and evolving human-AI partnerships, fully realizing the framework's potential.

# Discussion

The human-AI handshake framework offers a comprehensive approach to fostering dynamic, ethical, and adaptable partnerships between humans and AI systems. It emphasizes bi-directional collaboration, mutual reinforcement, and ethical safeguards, addressing key gaps in traditional human-centered AI (HCAI) models. By integrating principles such as transparency, adaptability, and user engagement, the framework proposes a shift toward AI systems that evolve alongside users, functioning as partners that augment human capabilities while ensuring ethical alignment.

Current AI tools demonstrate partial alignment with the framework, showcasing progress in facilitating bi-directional interaction and capability augmentation. For instance, GitHub Copilot and ChatGPT enhance productivity and creativity by enabling users to refine outputs interactively and iteratively. Similarly, tools like Adobe Firefly and DALL·E exemplify how AI can augment workflows in creative industries, blending generative capabilities with human originality. High-stakes systems, such as healthcare AI platforms and autonomous technologies like Tesla Autopilot, incorporate validation mechanisms to ensure human oversight. However, these tools often remain static in their learning capabilities and lack deep contextual adaptability, falling short of fostering the co-evolutionary dynamics that the framework envisions.

Despite advancements, significant challenges hinder the full implementation of the human-AI handshake framework. Limited dynamic learning is a major barrier, as most AI tools rely on static training models, restricting their ability to adapt to evolving user needs and contexts. This lack of mutual learning undermines the goal of bi-directional, co-evolutionary relationships. Additionally, transparency gaps are prevalent, as many AI systems operate as "black boxes," obscuring their decision-making processes and eroding user trust. Interpretable AI models are essential for fostering meaningful collaboration. Ethical concerns further constrain the application of the framework in sensitive domains like law, healthcare, and public policy, where biases in training data and insufficient fairness mechanisms can lead to misuse. Furthermore, many current AI tools are designed for task-specific purposes, limiting their flexibility and utility in broader, multi-domain applications.

Addressing these challenges can unlock transformative potential across diverse domains. In healthcare, AI systems could integrate clinical expertise with data-driven insights, enabling precise diagnostics and personalized patient care. Implementing bi-directional feedback loops would allow for dynamic refinement of treatment strategies. In education, adaptive learning platforms that incorporate mutual learning and bi-directional feedback could provide personalized and effective teaching strategies tailored to individual student needs. Within creative industries, co-creative AI tools could foster greater innovation by augmenting human creativity while preserving originality and agency. Finally, in public policy and law, transparent and ethically aligned AI systems could enhance policy analysis and legal research, promoting fairness, accountability, and inclusivity.

To fully realize the framework's potential, targeted advancements are necessary. First, advanced learning mechanisms such as reinforcement learning and federated learning should be employed to achieve real-time responsiveness and long-term adaptability. Transparency and explainability must also be prioritized by developing hybrid AI models that combine interpretable decision-making processes with deep learning architectures, enhancing user trust. Ethical safeguards are essential to ensure fairness, and should include mechanisms for real-time bias detection, fairness audits, and ethical oversight integrated into AI design and deployment. Moreover, AI systems should be designed to support mutual evolution, refining outputs continuously based on user feedback to foster reciprocal growth and alignment with human values. Finally, empirical validation through domain-specific studies is crucial to assess the framework's scalability, effectiveness, and alignment with real-world applications.

A structured research agenda is critical to refine and validate the human-AI handshake framework. Empirical studies involving interviews and surveys with AI practitioners, end-users, and stakeholders would provide valuable insights into practical challenges and opportunities for implementing the framework. Tool evaluation across a wider range of AI applications in various industries would further inform how the framework can be applied effectively in different contexts. Additionally, iterative testing of the framework in real-world scenarios will be key to collecting data on usability, adaptability, and ethical alignment, which can guide further refinements and improvements.

In conclusion, the human-AI handshake framework offers a visionary path for advancing human-AI collaboration, emphasizing transparency, adaptability, and ethical alignment. While current AI tools have made notable progress in aligning with some of these principles, addressing gaps in adaptability, transparency, and ethics is essential to achieving the framework's full vision. By prioritizing these areas and fostering continuous innovation, human-AI partnerships can be transformed into dynamic, trustworthy, and ethically sound collaborations that empower users and enhance outcomes across sectors. Future research should focus on empirical validation, diverse real-world applications, and scalability to refine the framework further. These efforts will deepen our understanding of human-AI collaboration, advancing the field and unlocking new possibilities for innovation while preserving human agency and ethical integrity.